\documentclass[12pt]{iopart}
\usepackage{epsf}
\begin{document}

\newcommand{\dpar}[1]{\frac{\partial}{\partial #1}} 
\newcommand{\grad}{\vec{\nabla}}

\title[Strangeness Production in Microscopic Transport Models]
{Strangeness Production in Microscopic Transport Models}

\author{Steffen A. Bass
\footnote[1]{bass@phy.duke.edu}
}

\address{Department of Physics, Duke University\\
	\& RIKEN-BNL Research Center, Brookhaven National Laboratory}

\begin{abstract}
Strangeness production in microscopic transport models for relativistic
heavy-ion collisions from SIS to RHIC is reviewed: after a brief 
introduction into elementary strangeness production processes, the
main emphasis is put on strangeness as indicator of the nuclear equation
of state, the excitation function of the $K^+/\pi^+$ ratio and strangeness
as a deconfinement indicator.
\end{abstract}




\section{Introduction and overview}

Microscopic transport models are a unique tool for the study of
relativistic heavy-ion collisions: 
they offer a method
of connecting the observable final state of such a collision
with the time-evolution of the reaction, its dynamics and many not directly
observable but very much sought after quantities and phenomena like the nuclear
equation of state and the deconfinement phase-transition to  
a Quark-Gluon-Plasma \cite{reviews}. 
Progress in heavy-ion physics requires a symbiosis
between experiment and (transport-)theory.

The basic principles of microscopic transport models can best be explained
by studying the Vlasov-Uehling-Uhlenbeck (VUU) equation:

\begin{eqnarray*}
\left[ \dpar{t}  \right. & +& \left. 
\left( \frac{{\bf p}_1}{m}+\grad_{p_1}\Re\Sigma^+ \right) \cdot \grad_{r_1} -
\grad_{r_1}\Re \Sigma^+ \cdot\grad_{p_1} \right] 
f_1({\bf r}_1,{\bf p}_1,t)
\nonumber \\
& = & 
\frac{2g}{m^2 (2\pi \hbar)^3}\int {\rm d}^3 {\bf p}_2 \int {\rm d}^3
{\bf p}'_1
\int {\rm d}^3 {\bf p}'_2 \, \delta^4(p_1+p_2-p'_1-p'_2)
 \frac{{\rm d}\sigma}{{\rm d}\Omega}
\nonumber \\
&& 
\times \left[ f'_1 f'_2 { (1-f_1)(1-f_2)} -
 f_1 f_2 { (1-f'_1)(1-f'_2)} \right]  
\end{eqnarray*}

This equation describes the time-evolution of the one-particle distribution
function $f_1({\bf r}_1,{\bf p}_1,t)$ which contains the positions and
momenta of all particles of the system 
(i.e. the microscopic degrees of freedom) -- initially all the protons
and neutrons  (or, in the case of a partonic
description, the quarks and gluons) of the two colliding nuclei.
These particles may interact either through an interaction
given by the the real part of the retarded self-energy $\Re \Sigma^+$ 
(often also approximated by and referred to as the mean field) or
through binary scattering, symbolized in this equation by
the differential cross section $\frac{{\rm d}\sigma}{{\rm d}\Omega}$.
If the r.h.s of this equation, the collision
integral, is neglected one obtains the Vlasov equation. On the
other hand, if the real part of the retarded self-energy $\Re \Sigma^+$ is
neglected and if the so called Pauli blocking factors $(1-f)$ in the
collision integral are approximated by 1, one obtains the famous Boltzmann
equation.

\begin{figure}
\begin{center}
\epsfxsize=0.7\textwidth\epsfbox{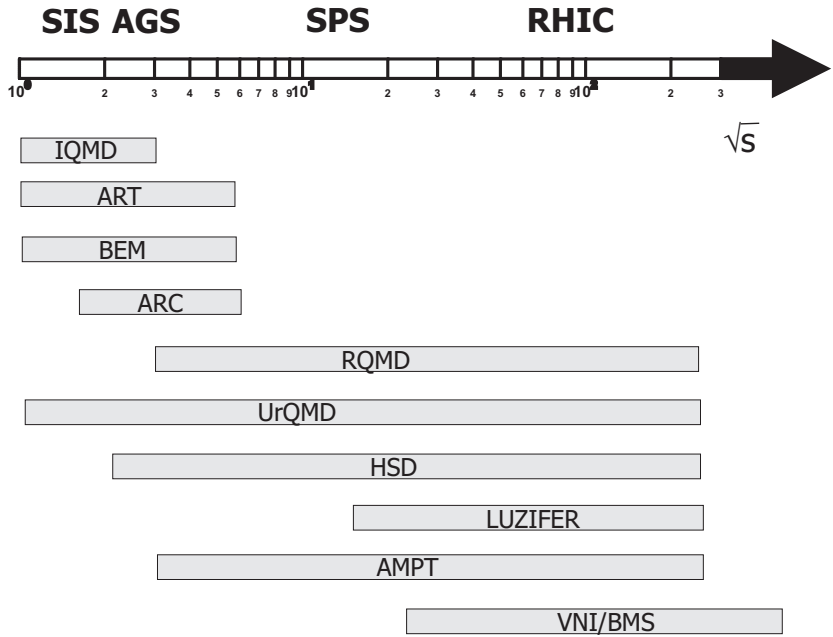}
\end{center}
\caption{\label{micmods} Overview of available microscopic transport
models and their range of applicability in incident beam energy.}
\end{figure}

In terms of the details of the implementation of mean field and collision
term (e.g. in the number of hadronic resonances included or in the 
parametrization of cross sections), 
transport models may vary widely. Figure~\ref{micmods} provides
an overview of currently available microscopic transport models applicable
from SIS up to RHIC energies: models with purely hadronic degrees of freedom
utilizing both, a mean field as well as a collision term are 
(I)QMD \cite{iqmd}, VUU/BUU \cite{buu}, ART \cite{art} and BEM \cite{bem}.
Models not containing a mean field are commonly referred to as cascade
models -- ARC \cite{arc} and LUZIFER \cite{luzifer} fall into this category. 
At higher energies
initial particle production requires the introduction of
string excitations -- models which combine strings and hadrons are
RQMD \cite{rqmd}, UrQMD \cite{urqmd} and HSD \cite{hsd}. At RHIC energies
strings and hadrons may not anymore be the relevant degrees of freedom --
here deconfinement needs to be taken into account and the elementary
degrees of freedom for the initial reaction stage are quarks and gluons
interacting through hard scattering as in VNI/BMS \cite{bms} and AMPT 
\cite{ampt}. 

\section{Mechanisms of strangeness production}

Strangeness may be produced either in initial collisions among
the incoming nucleons
of the two colliding nuclei or through secondary interactions among 
produced particles, e.g. pions and nucleons or excited resonances.

\subsection{Strangeness production at threshold}

\begin{figure}
\centerline{\epsfxsize=0.6\textwidth\epsfbox{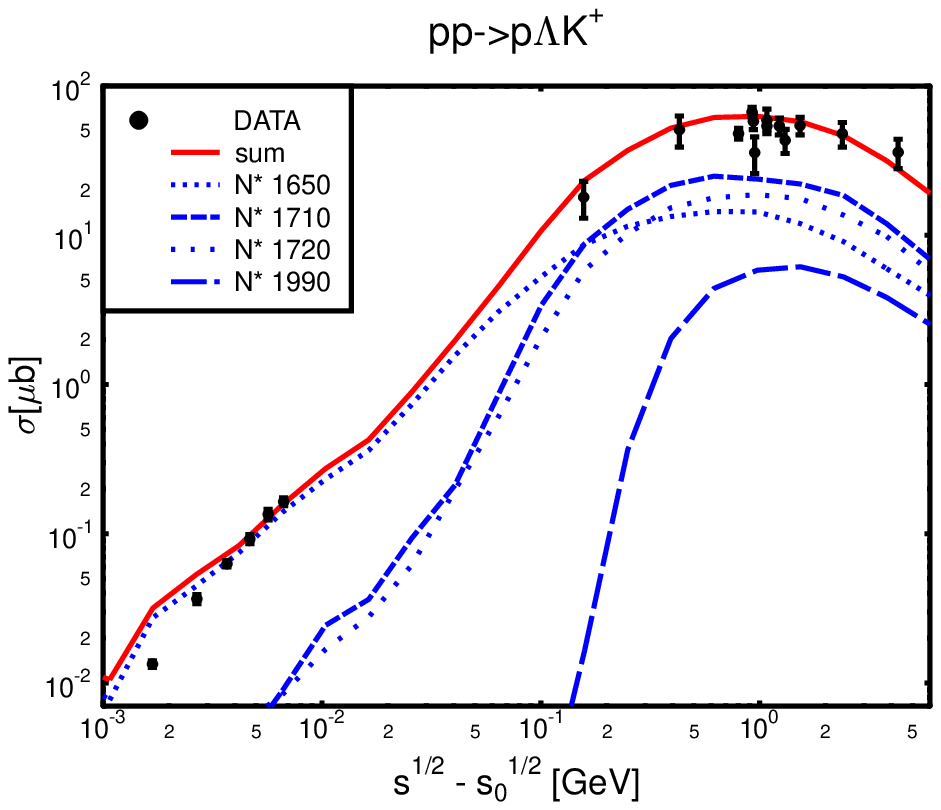} 
\hfill \epsfxsize=0.4\textwidth\epsfbox{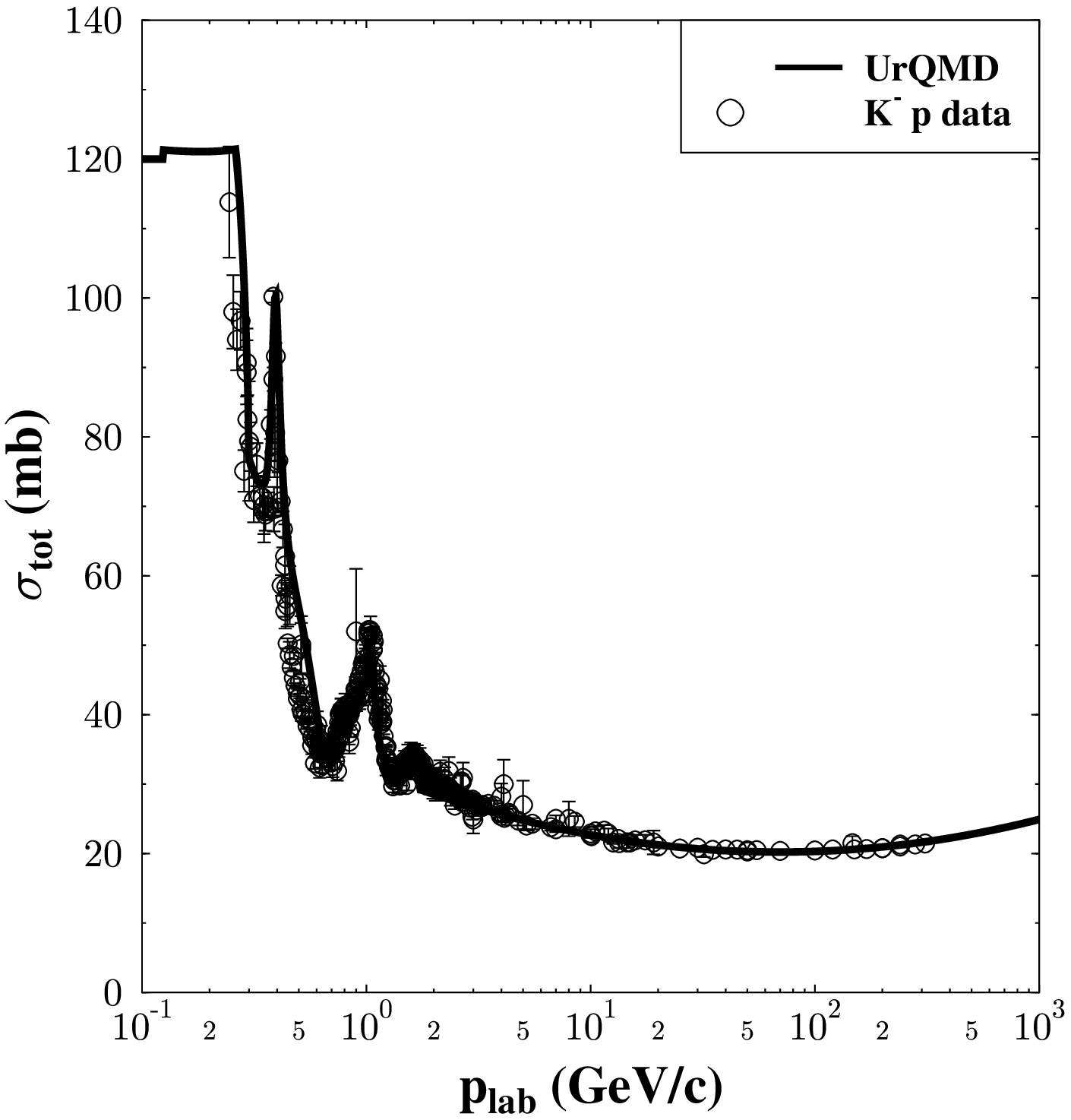}}
\caption{\label{elementary1} Left: excitation function of the 
exclusive $p+p \to p+\Lambda+K^+$ production cross section close 
to threshold in a resonance model. Right: $K^- + p$ scattering cross
section in UrQMD compared to data.}
\end{figure}

There are two different approaches for initial strangeness production
close to threshold:
\begin{enumerate}
\item {\em direct parametrization}: the individual cross sections
	for different exclusive production channels, e.g. 
	$p+p \to p+\Lambda+K^+$ are directly parameterized. This method
	is very accurate if the cross sections are well known (i.e. well
	measured) but may become cumbersome away from threshold with
	increasing number of explicit channels to parameterize. 
\item {\em resonance model}: here strangeness production is a two step
	process. Initially a heavy baryon resonance is excited, e.g. via
	$p+p\to N + N^*_{1710}$, which subsequently decays via
	$ N^*_{1710} \to Y+K^+$. This approach allows for an easier 
	extension into the higher energy domain and may provide some
	rudimentary guidance for unknown strangeness production cross
	sections in secondary collisions,
	e.g. $\pi+N\to N^*_{1710} \to Y+K^+$. The left frame of 
	figure~\ref{elementary1} shows a fit of the resonance model to
	the exclusive $p+p \to p+\Lambda+K^+$ reaction channel.
\end{enumerate}

Secondary interactions like pion-induced strangeness production
or flavor-exchange reactions are at least as important for the reaction
dynamics and final strangeness yield as the initial/primordial strangeness
production channels. The right frame of figure~\ref{elementary1} shows
the $K^-+N$ reaction cross section, which exhibits a distinct resonance
structure. The hyperon resonances which are excited via this cross section
may then either decay again into the $K^-+N$ channel, or to almost equal
probability decay into the $Y+\pi$ channel, thus transferring strangeness
in and out of baryonic degrees of freedom. As we shall later see, this
$K^-+N \leftrightarrow Y+\pi $ exchange reaction is of particular importance.

Ambiguities in both types of approaches arise from unknown and mostly
immeasurable cross sections such as strangeness production in interactions
involving baryon- and meson-resonances, e.g. $\Delta_{1232} + N \to K^+ + X$.

\subsection{Strangeness production at high energies}

\begin{figure}
 \centerline{\epsfxsize=0.5\textwidth\epsfbox{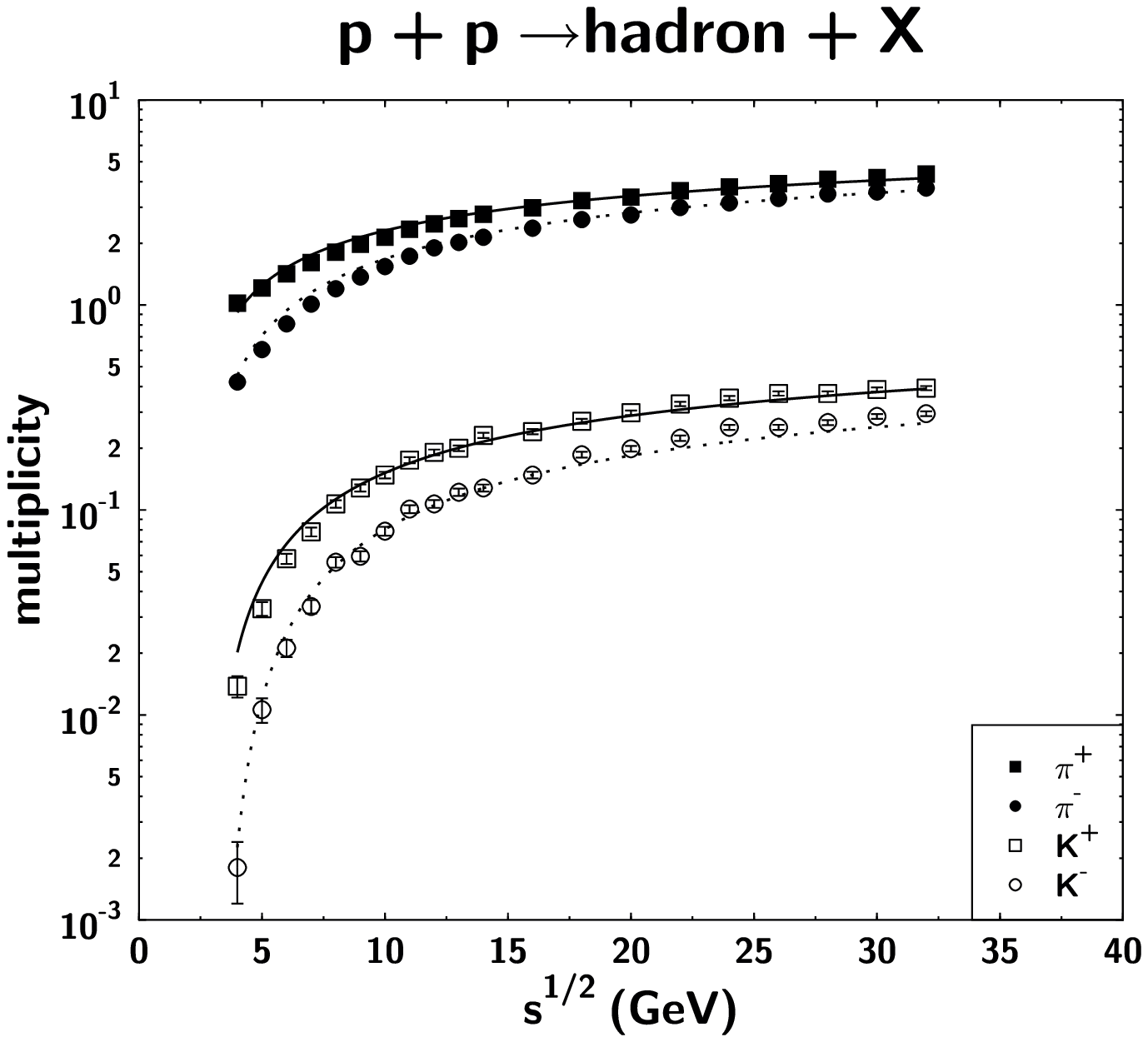} 
\hfill \epsfxsize=0.52\textwidth\epsfbox{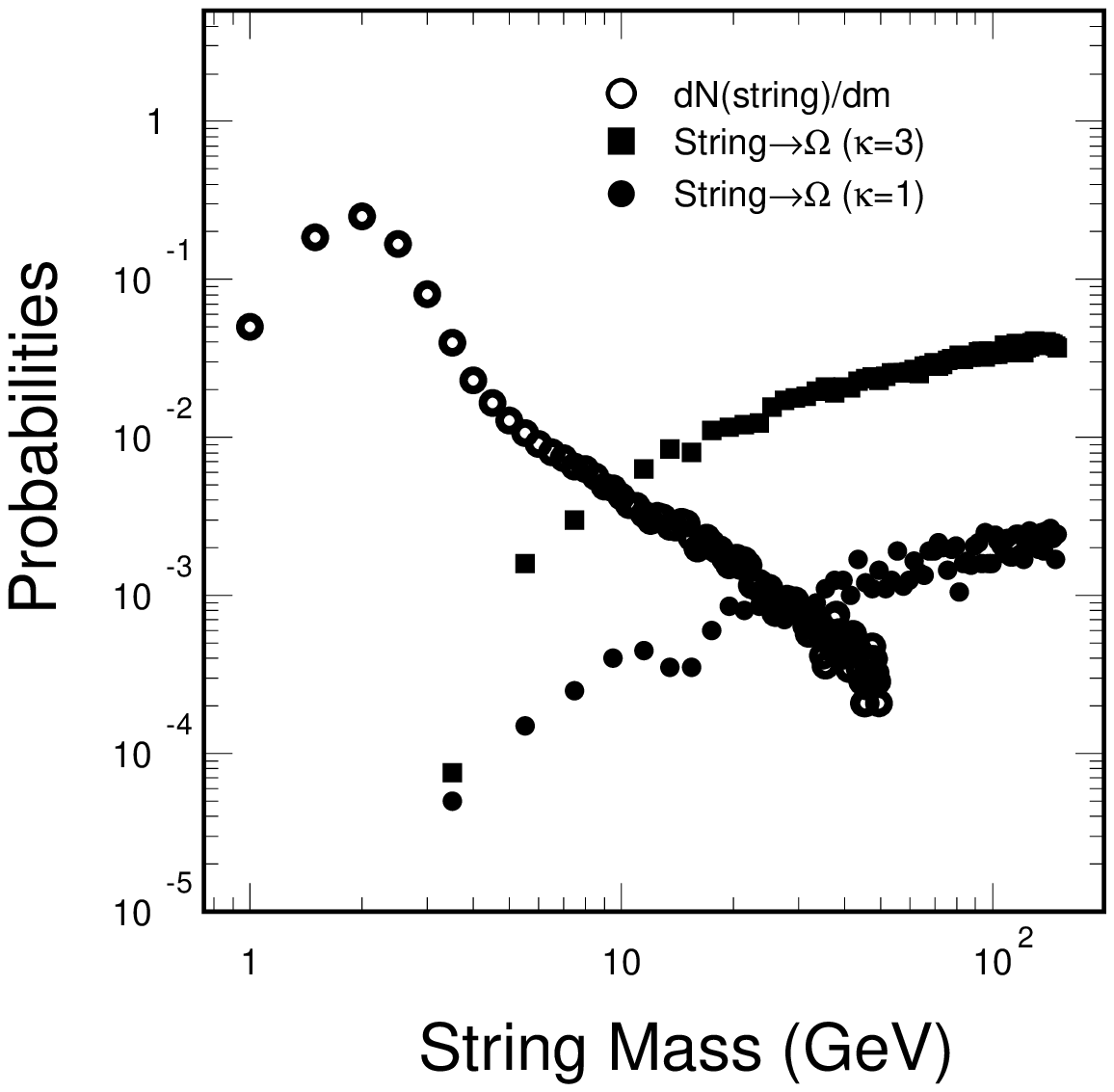}}
\caption{\label{elementary2} Left: excitation function of the $K$ and $\pi$ 
multiplicity in elementary $p+p$ reactions in UrQMD at large CM-energies
compared to data. Right: distribution of string masses in Pb+Pb reactions
at the CERN-SPS (160 GeV/u) and the $\Omega^-$ production probability 
as a function of string-mass for two different values of the string-tension
$\kappa$.
}
\end{figure}

At higher incident beam energies, particle production in general is dominated
by string excitation and fragmentation. Up to 70\% of the total strangeness
produced in a Pb+Pb collision at top CERN-SPS energies is produced in initial
highly energetic nucleon-nucleon interactions which lead to the excitation
and subsequent fragmentation of strings \cite{urqmd}. On an elementary
hadron-hadron level, the parameters of the string fragmentation
are fitted to measured multiplicities and momentum distributions. The
left frame of figure~\ref{elementary2} shows the resulting excitation function
of the pion and kaon multiplicity in proton-proton reactions in the UrQMD
model (plot symbols are data, lines the model fit). 

When going from elementary hadron-hadron interactions to collisions of heavy
nuclei one can explore the possibility of medium effects due to the
hot and dense environment of the reaction. One such (non-hadronic) medium 
effect is the formation of color-ropes \cite{ropes} -- overlapping strings
form a chromo-electric field which due to its larger field strength
compared to the individual color-fluxtubes has an enhanced probability
of fragmenting into strange hadrons. Rope effects can be simulated by
increasing the string-tension $\kappa$ from it's vacuum value of 
$\kappa=1$~GeV/fm to a value of $\kappa=3$~GeV/fm for strings fragmenting
in the high energy-density region of the collision. The right frame of
figure~\ref{elementary2} shows the distribution of string masses excited
in a Pb+Pb collision at 160 GeV/u as well as the fragmentation probability
of the string into an $\Omega^-$, both for $\kappa=1$~GeV/fm as well
as for $\kappa=3$~GeV/fm \cite{blubb1}. 
A drastic enhancement of the $\Omega^-$ formation
probability by more than one order of magnitude is visible.

\section{Strangeness and the nuclear equation of state}

\begin{figure}
 \centerline{\epsfxsize=0.45\textwidth\epsfbox{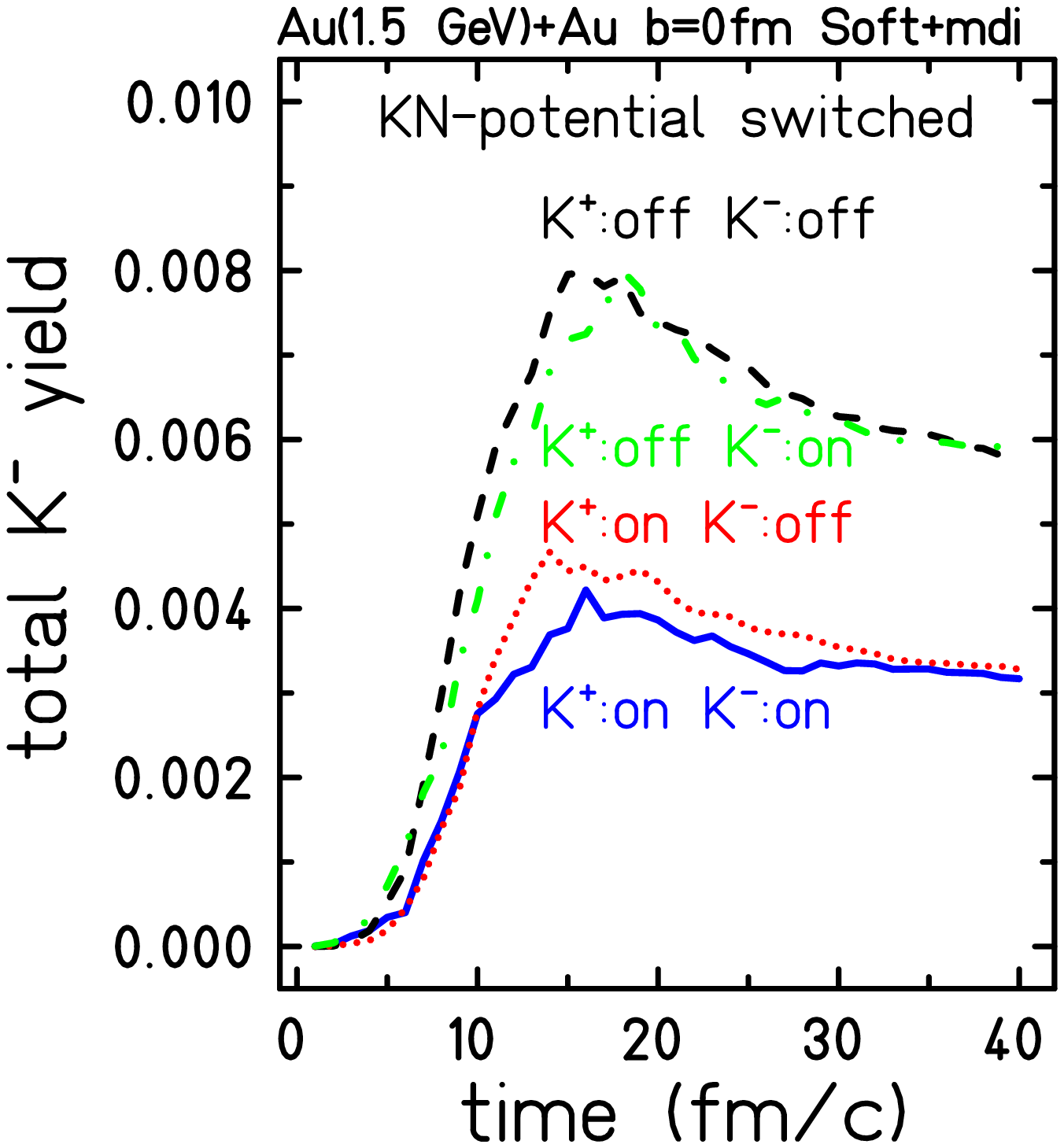} 
\hfill \epsfxsize=0.35\textwidth\epsfbox{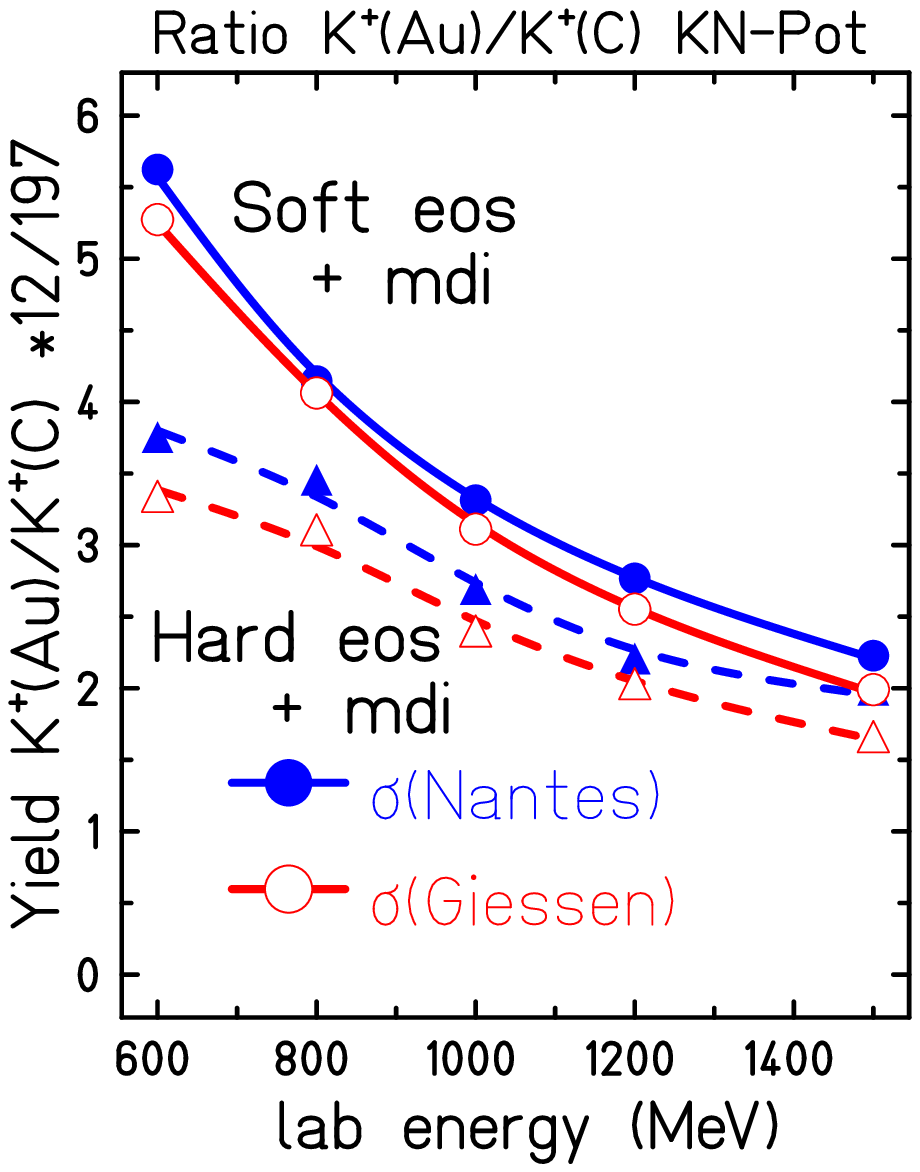}}
\caption{\label{sis} Left: time-evolution of the $K^-$ yield with and
without inclusion of $K^\pm$ potentials. Right: ratio of energy-spectra
of $K^+$ in Au+Au and C+C collisions for different cross section 
and equation of state parametrizations.
}
\end{figure} 

In the SIS and AGS energy domain the main issues to be addressed by 
microscopic transport theory with respect to strangeness are the existence
of in-medium modifications for kaons and the nuclear equation of state.
In particular the $K^-$ is considered to be strongly influenced by
a scalar as well as a vector interaction. The left frame of figure~\ref{sis}
shows the time-evolution of the $K^-$ yield for different combinations
of potentials acting upon the $K^-$ or the $K^+$ \cite{hartnack1}. 
Surprisingly the $K^-$
yield is totally insensitive to $K^-$ potentials but very sensitive to
$K^+$ potentials. This flavor coupling can be well understood through
the flavor-exchange reaction $K^-+N \leftrightarrow Y+\pi$ with the
the hyperons being linked to the $K^+$ and their potential 
via associated production or higher
resonance states like $ N^*_{1710} \leftrightarrow Y+K^+$.
The main observable to explore potential effects on $K^-$ is the so called
kaon-flow, which has been shown to be extremely sensitive to the strength
and nature of the real part of the kaon-nucleon interaction \cite{k_flow}.

The $K^+$ yield, on the other hand, has been found to be very sensitive
to the nuclear equation of state \cite{k_eos}. Ambiguities may arise
due to different implementations of unknown production cross sections like
$\Delta_{1232}+N \to N+Y+K^+$. In order to avoid these, ratios of $K^+$ 
spectra for heavy and light collision systems can be used \cite{k_ratio}.
The right frame of figure~\ref{sis} shows the ratio of the $K^+$ energy
spectra for Au+Au and C+C collisions for different cross section 
and equation of state parametrizations \cite{hartnack2}. 
A strong sensitivity to the
equation of state is observed while different production cross sections
do not alter the value or functional form of the ratio. Comparisons of 
current calculations with data favor a soft equation of state with 
momentum-dependent interaction \cite{k_ratio}.

\section{Excitation functions}

\begin{figure}
 \centerline{\epsfxsize=0.4\textwidth\epsfbox{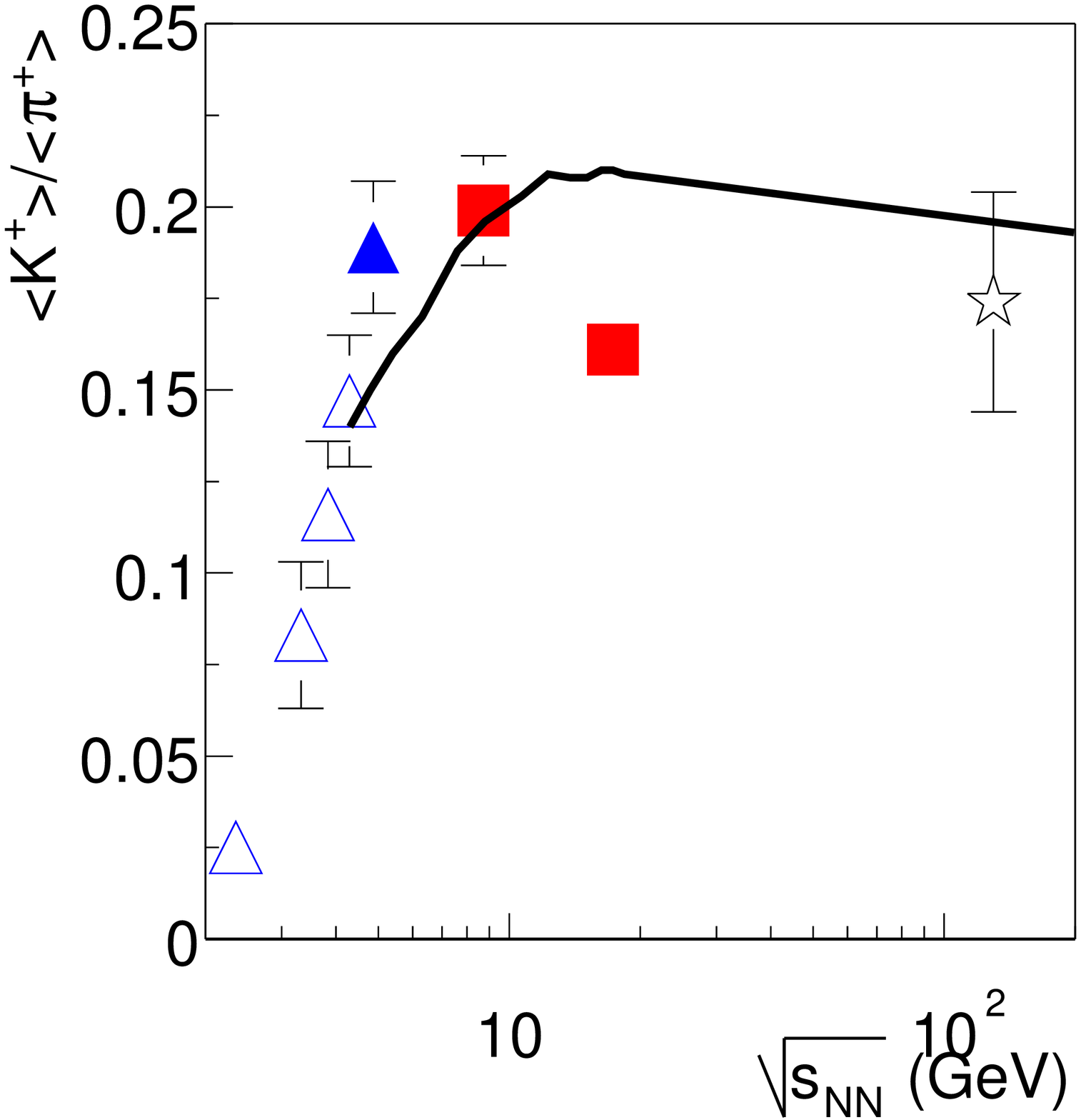} 
\hfill \epsfxsize=0.55\textwidth\epsfbox{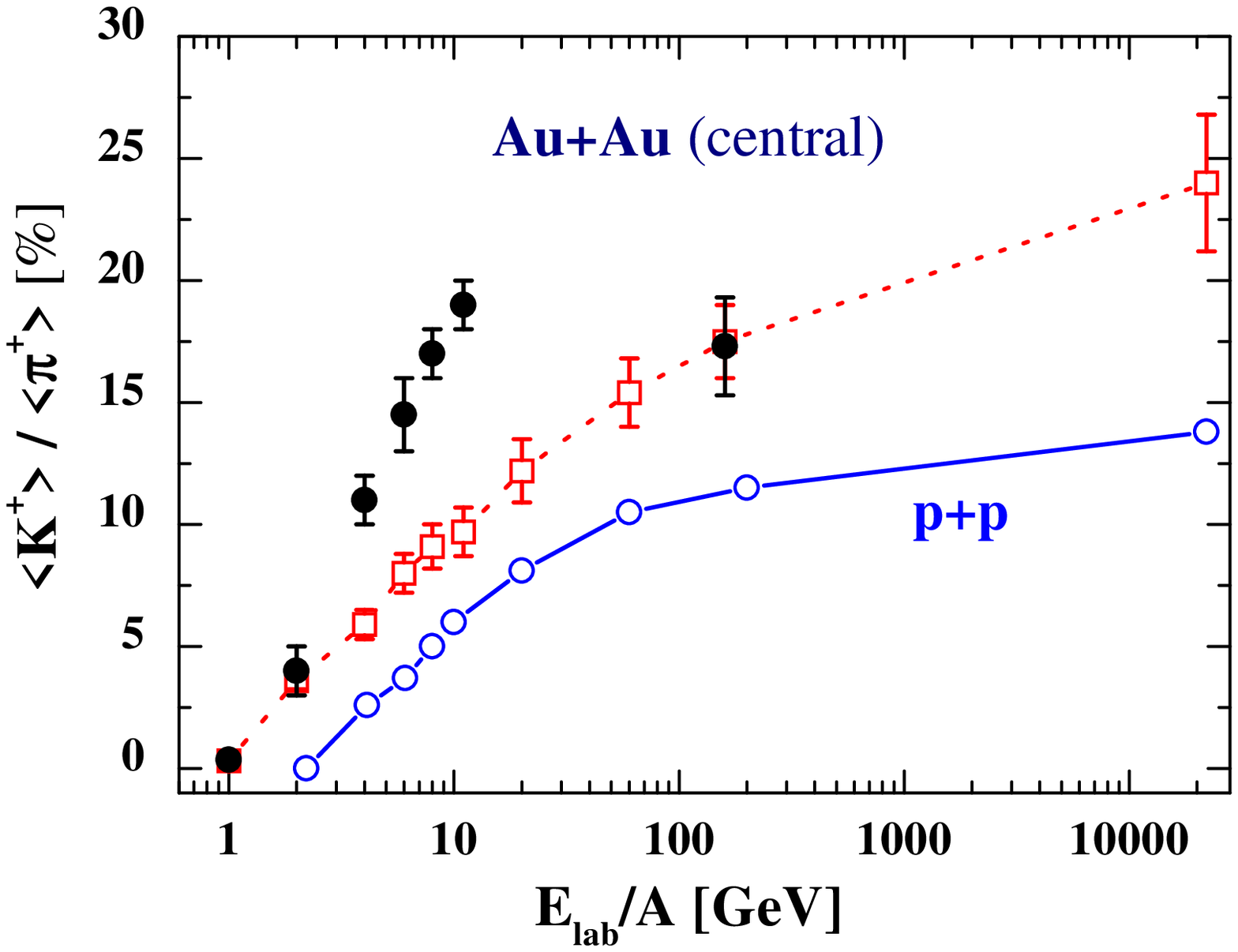}
}
\caption{\label{kpi1} Excitation function of the $K^+/\pi^+$ ratio in
RQMD (left) and HSD (right; open symbols refer to HSD, closed symbols
are data). 
}
\end{figure} 

The $K^+/\pi^+$ ratio provides a measure of the ratio of newly produced
strange to non-strange valence-quarks in a heavy-ion reaction. A kink
in its excitation function vs. incident beam energy was thought to hint
at a possible deconfinement phase-transition. Recent calculations
in the framework of a statistical model show that the
$K^+/\pi^+$ ratio is expected to reach its maximum value around a beam
energy of 30 GeV/u, well in line with experimental findings \cite{sm}. 
While the
agreement of the data with statistical model calculations cannot per se prove
or disprove the existence of a deconfinement phase-transition, the $K^+/\pi^+$
ratio as deconfinement indicator remains ambiguous at best.

\begin{figure}
 \centerline{\epsfxsize=0.48\textwidth\epsfbox{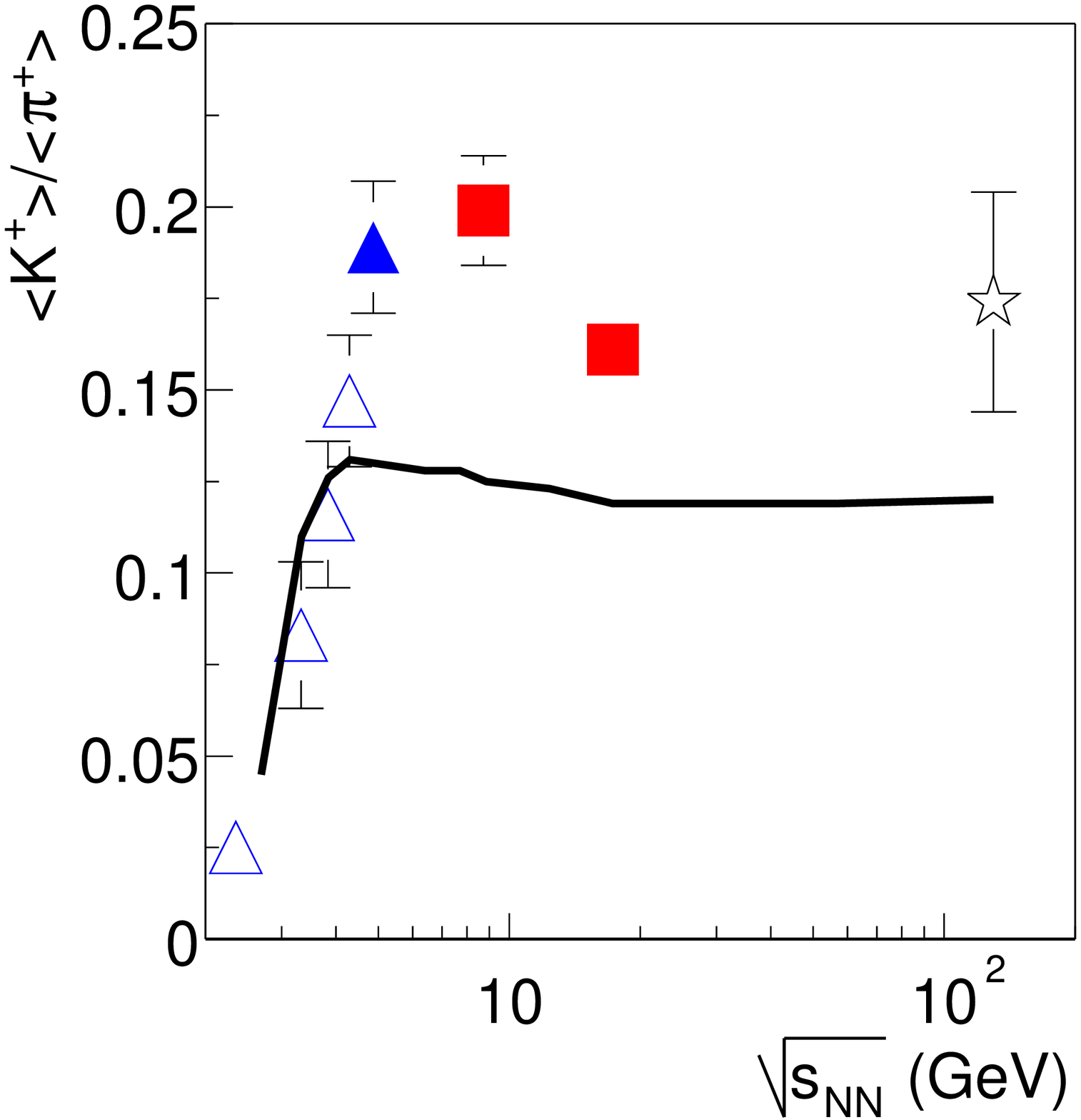} 
\hfill \epsfxsize=0.52\textwidth\epsfbox{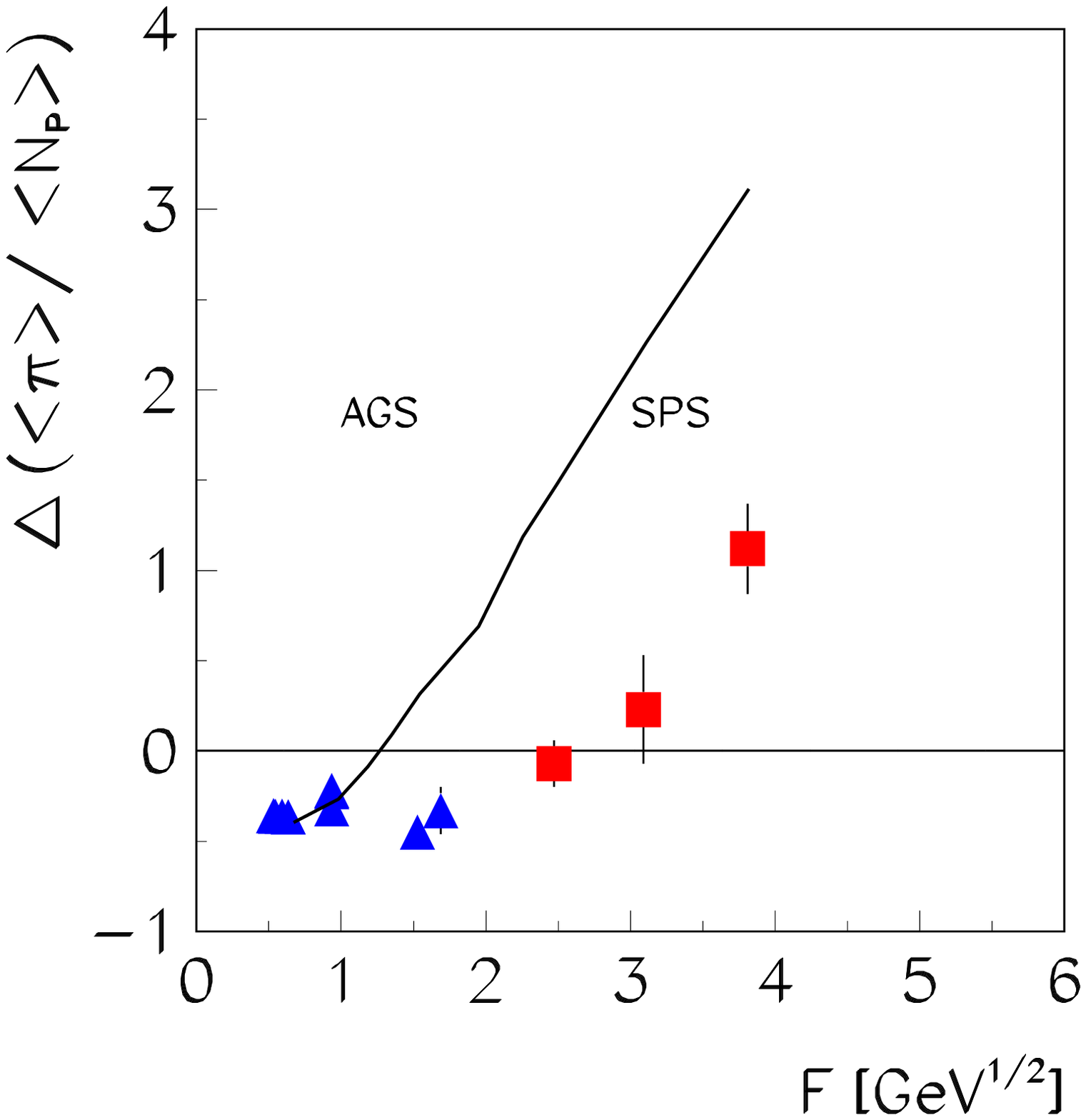}
}
\caption{\label{kpi2} Left: excitation function of the  $K^+/\pi^+$ in UrQMD.
Right: comparison of the pion per participant excess in Pb+Pb reactions 
vs. p+p reactions in UrQMD to data.

}
\end{figure}

However, the beam-energy dependence of the  $K^+/\pi^+$ ratio has emerged as
a rather stringent test for microscopic transport models: figures~\ref{kpi1}
and~\ref{kpi2} show the excitation function of the $K^+/\pi^+$ ratio for
RQMD (fig.~\ref{kpi1} left, provided by \cite{marek}), 
HSD (fig~\ref{kpi1} right, taken from \cite{hsd_kpi}) and UrQMD 
(fig.~\ref{kpi2} left, provided by \cite{henning,marek}).  
While RQMD compares well to the data, both HSD and UrQMD exhibit problems
which may not necessarily be rooted explicitly in the strangeness production
mechanisms of the particular models: in the case of HSD, an analysis shows
that the observed functional dependence of the $K^+/\pi^+$ ratio is 
very sensitive to the threshold of the string production cross section and
the excitation of the high-mass resonance continuum \cite{carsten}. Older
versions of RQMD with a similar implementation of these cross sections 
exhibit the same behavior as observed in HSD.
In the case of UrQMD the under-prediction of the $K^+/\pi^+$ ratio starting at
AGS energies can be traced to a pion excess (the kaon yields roughly agree
with the data). The right frame of figure~\ref{kpi2} shows
the pion per participant excess in Pb+Pb reactions 
vs. p+p reactions in UrQMD compared to data \cite{henning,marek}: in UrQMD 
this excess sets in too early with respect to the incident beam energy.
It is very likely that this behavior is caused by the 
violation of detailed balance due to the lack of
multi-particle (mostly pion) collisions which would convert the pion excess
into heavier particles such as anti-baryons and hyperons \cite{rapp,greiner}
-- preliminary calculations incorporating these processes yield a
$K^+/\pi^+$ ratio very close to the data \cite{henning}. 

A more detailed study of various excitation functions of 
the $K^+/\pi^+$ ratio in RQMD as well as a discussion
of the $K^+/K^-$ ratio can be found in \cite{rqmd_kaons}.

\section{Strangeness as deconfinement indicator}

\begin{figure}
 \centerline{\epsfxsize=0.52\textwidth\epsfbox{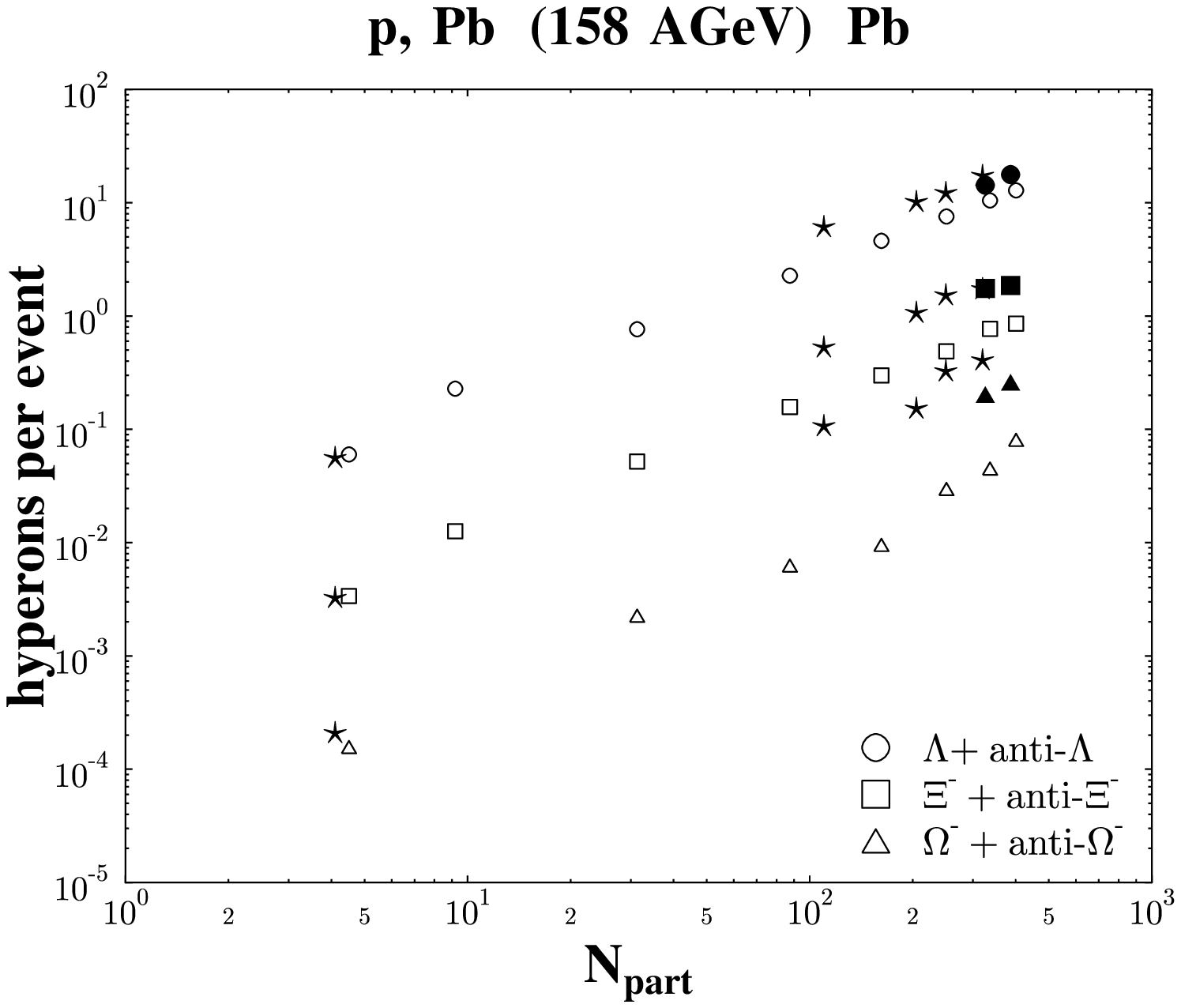} 
\hfill \epsfxsize=0.48\textwidth\epsfbox{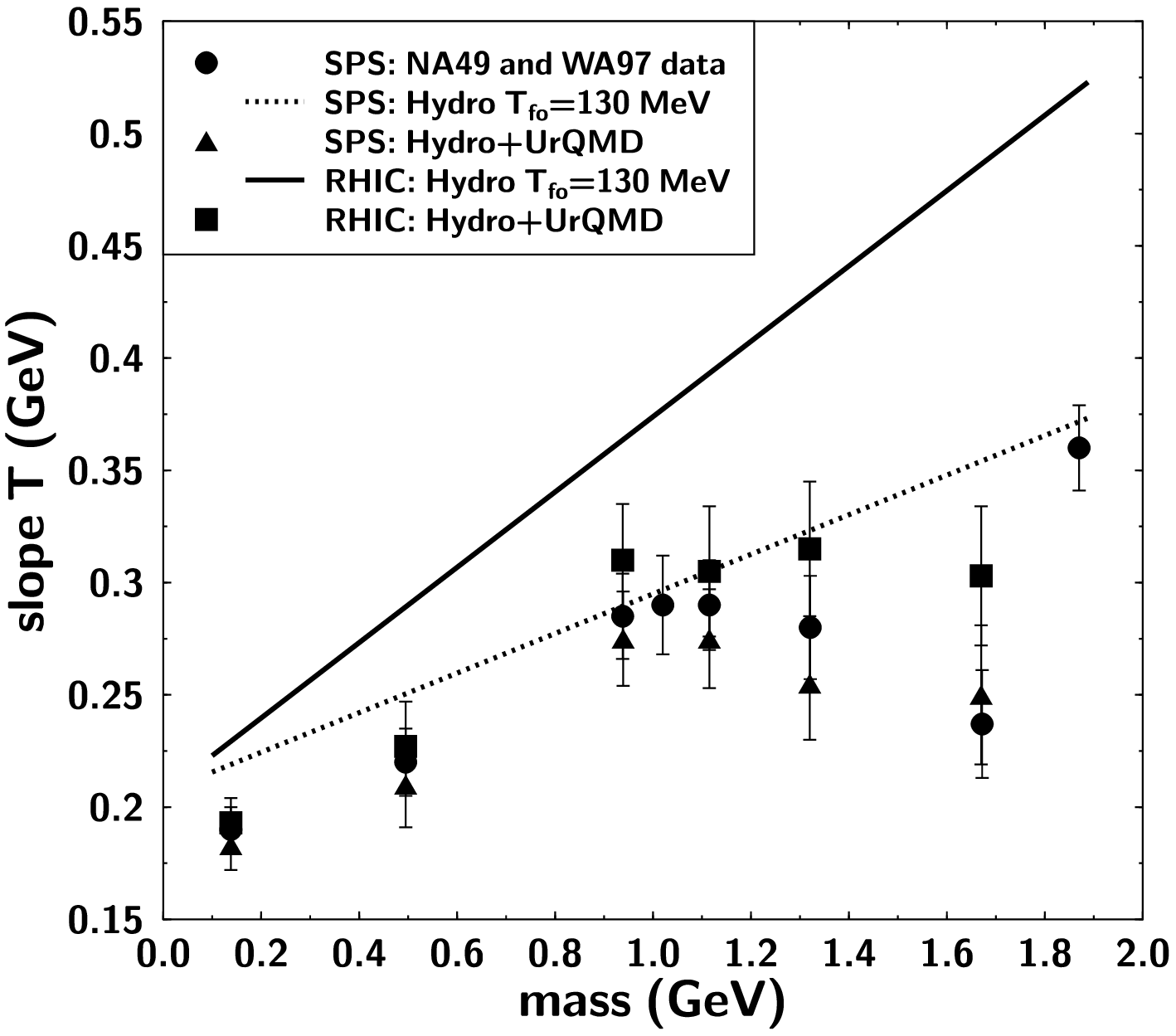}}
\caption{\label{sps} Left: excitation function of strange baryon multiplicity
vs. number of participants in UrQMD. Right: Inverse slopes of the 
$m_T$-spectra of $\pi$, $K$, $p$,
$\Lambda+\Sigma^0$, $\Xi^0+\Xi^-$, and $\Omega^-$ at $y_{c.m.}=0$. 
}
\end{figure}

The relative enhancement of strange and especially multistrange baryons 
with respect to peripheral (or proton induced) 
interactions has been suggested as a signature for the transient 
existence of a QGP-phase \cite{raf8286,koch86,koch88}:  
the main argument being that the (chemical or flavor) equilibration 
times should be much shorter in the plasma phase than in 
a thermally equilibrated hadronic fireball of $T\sim 160\,$MeV.

The dominant production mechanism in an equilibrated (gluon rich) plasma 
phase,  namely the production of $s\overline{s}$ pairs via gluon fusion 
($gg \rightarrow s \overline{s}$) \cite{raf8286},  
should allow for equilibration times similar to the 
interaction time of the colliding nuclei, and to the expected plasma lifetime 
(a few fm/c).

The yields of strange baryons per event calculated in UrQMD 
are shown in the left frame of figure~\ref{sps} 
as a function of the number 
of participants  for Pb+Pb and p+Pb collisions at 160 GeV/u \cite{sven}.
The $\Lambda+\overline{\Lambda}$- (circles), 
$\Xi^- + \overline{\Xi^-}$- (squares), 
and $\Omega^- + \overline{\Omega^-}$- (triangles) values 
are shown.
The stars correspond to experimental data of the 
WA97 collaboration \cite{strange_data}. Open symbols represent the results 
of the standard UrQMD calculations, whereas 
full symbols exhibit a calculation 
with an enhanced string tension of $\kappa=3\,$GeV/fm, 
for the most central collisions ($N_{\rm part}\ge 300$).
Obviously the standard UrQMD calculation, which can be 
seen as a {\em baseline} of known hadronic physics, strongly underestimates
the (multi-)strange particle yields in central collisions, in particular in
the case of the $\Omega^-$. Only the inclusion of non-hadronic medium effects,
like color-ropes \cite{ropes}, which are simulated by increasing the
string-tension for central collisions (see also the right frame
of figure~\ref{elementary2}), enhances the yield to a level of 
near-compatibility with the data. Similar findings have also been made
in the context of HIJING calculations \cite{hijing}. While these findings
by no means prove the validity of the color-rope approach, they clearly show
the necessity of some kind medium effect beyond regular binary hadronic
(re)scattering in order to understand the data. 
This statement is corroborated by a calculation with the string model NeXuS,
including so-called QGP droplets -- domains of high energy-density
hadronizing according to a statistical phase-space population -- which
give rise to a similar strangeness enhancement \cite{nexus}.

However, recently 
hadronic multi-particle interactions in the early dense reaction phase have
been suggested to significantly enhance the yield of anti-protons and 
(anti-)hyperons \cite{rapp,greiner}. It remains to be seen, however, whether
these effects are sufficient to explain the observed $\Omega^-$ enhancement
or other non-hadronic (i.e. deconfinement based) effects need to be 
taken into account.

The study of deconfinement and a subsequent phase-transition to 
deconfined hadronic matter poses a great challenge to microscopic transport
models. In most approaches hadronization is a uni-directional process
and the equation of state of the system ill-defined. One possible remedy 
is to use a hydrodynamical approach for the early deconfined phase of the 
reaction and subsequent phase-transition, coupled with a microscopic 
calculation for the later, hadronic, reaction phase in which the hydrodynamical
assumptions are not valid any longer \cite{hu}. In the following,
such a combined macro+micro model will be used to study the flavor- and 
mass-dependence of hadronic
slope parameters: the right frame of fig.~\ref{sps}
displays these inverse slope parameters $T$ 
obtained by an exponential fit to
$dN_i/d^2m_Tdy$ in the range $m_T-m_i<1$~GeV for SPS and RHIC 
\cite{dumi1} and
compares them to SPS data \cite{strange_data}.
The trend of the data, namely the ``softer'' spectra of $\Xi$'s and
$\Omega$'s as compared to a linear $T(m)$ relation is reproduced
reasonably well. This is in contrast to ``pure'' hydrodynamics with kinetic
freeze-out on a common hypersurface (e.g.\ the $T=130$~MeV isotherm), where
the stiffness of the spectra increases linearly with mass
as denoted by the lines in fig.~\ref{sps}.
When going from SPS to RHIC energy, such a hybrid model as discussed here 
generally
yields only a slight increase of the inverse slopes,
although the specific entropy is larger by a factor of 4-5~!
The reason for this behavior is the first-order phase transition that
softens the transverse expansion considerably.       

The reason for the softening of the spectra is that
the hadron gas emerging from the hadronization of the QGP
is almost ``transparent'' for the multiple strange baryons.
Analyzing the collision numbers, one finds that
$\Omega$'s suffer on average only one hadronic interaction, whereas
$N$'s and $\Lambda$'s suffer far more
collisions with other hadrons before they freeze-out.
Thus, one may conclude that
the spectra of $\Xi$'s and especially $\Omega$'s  are practically
unaffected by the hadronic reaction stage and closely resemble 
those on the phase
boundary. They therefore act as probes of the QGP expansion prior
to hadronization and can be used to measure the expansion rate
of the deconfined phase.

\ack
I wish to thank M. Bleicher, R. Bramm, W. Cassing, J. Cleymans, A. Dumitru,
C. Fuchs, M. Gazdzicki, C. Greiner, C. Hartnack, C.M. Ko, T. Kolleger, 
S. Pal, S. Soff, S. Vance, H. Weber and N. Xu for many helpful discussions
and their support during the preparation of this review.
This work was supported by  RIKEN, Brookhaven
National Laboratory and DOE grants DE-FG02-96ER40945 as well as
DE-AC02-98CH10886.

\end{document}